# Current-induced magnetization switching in CoTb amorphous single layer

R. Q. Zhang[1], L. Y. Liao[1], X. Z. Chen[1], T. Xu[2], L. Cai[2], M. H. Guo[2], Hao Bai[1], L. Sun[3], F. H. Xue[3], J. Su[4], X. Wang[4], C. H. Wan[4], Hua Bai[1], Y. X. Song[1], R. Y. Chen[1], N. Chen[1], W. J. Jiang[2], X. F. Kou[3], J. W. Cai[4], H. Q. Wu[5], F. Pan[1] and C. Song[1,*]

[1]*Key Laboratory of Advanced Materials (MOE), School of Materials Science and Engineering, Tsinghua University, Beijing 100084, China*

[2]*State Key Laboratory of Low-Dimensional Quantum Physics and Department of Physics, Tsinghua University, Beijing 100084, China*

[3]*School of Information Science and Technology, ShanghaiTech University, Shanghai 201210, China*

[4]*Beijing National Laboratory for Condensed Matter Physics, Institute of Physics, Chinese Academy of Sciences, Beijing 100190, China.*

[5]*Institute of Microelectronics, Tsinghua University, Beijing 100084, China*

**We demonstrate spin-orbit torque (SOT) switching of amorphous CoTb single layer films with perpendicular magnetic anisotropy (PMA). The switching sustains even the film thickness is above 10 nm, where the critical switching current density keeps almost constant. Without the need of overcoming the strong interfacial Dzyaloshinskii-Moriya interaction caused by the heavy metal, a quite low assistant field of ~20 Oe is sufficient to realize the fully switching. The SOT effective field decreases and undergoes a sign change with the decrease of the Tb-concentration, implying that a combination of the spin Hall effect from both Co and Tb as well as an asymmetric spin current absorption accounts for the SOT switching mechanism. Our findings would advance the use of magnetic materials with bulk PMA for energy-efficient and thermal-stable non-volatile**

---

[*]songcheng@mail.tsinghua.edu.cn

**memories, and add a different dimension for understanding the ordering and asymmetry in amorphous thin films.**

Current-induced magnetization switching based on spin-orbit torque (SOT) in heavy metal/ferromagnet has great potential in magnetic random access memory (MRAM) [1−4], where an ultrathin CoFeB layer (~1 nm) with perpendicular magnetic anisotropy (PMA) [5] has been widely adopted. However, the thermal stability during device miniaturization and the film uniformity during wafer upgrading for the 1 nm-thick ferromagnetic functional layer remain challenging. Meanwhile, the PMA cannot preserve with increasing the ferromagnetic layer thickness, and the effective switching will be lost because of a limited spin coherence length [6,7] and seriously increased current density [8,9]. Beyond the traditional heavy metal/ferromagnet scenario, SOT switching in materials with global [e.g., (Ga,Mn)As] [10,11] or local (e.g., CuMnAs [12] and $Mn_2Au$ [13,14]) broken inversion symmetry has been realized. Such SOT has a bulk characteristic, which can solve the contradiction between high thermal stability and low critical switching current density. However, the need of low temperature for (Ga,Mn)As and the difficulty of robust signal readout for antiferromagnetic CuMnAs and $Mn_2Au$ limit their practical applications.

The structural symmetry breaking is a necessary condition for the SOT-induced deterministic magnetization switching [15]. Intuitively, one would consider that the superlattices and amorphous single layer films, which are designed to be symmetric, do not break the inversion symmetry. Nevertheless some recent works provide new perspectives for this issue. Bulk Dzyaloshinskii-Moriya interaction (DMI) [16] and chiral domain walls [17] were observed in amorphous GdFeCo single layer, which was attributed to the nonhomogeneous depth distribution of the rare earth within the film depth [16,18]. In symmetric epitaxial $[Co/Pd(111)]_N$ superlattices, a strong DMI with bulk character was reported [19], where the unequal strain of bottom Pd/Co and top Co/Pd interfaces [19,20] was explained to be the origin. These works seem to provide clues that thin films are always imperfect and symmetry broken exists everywhere. Then combining the demand for the high thermal stability in MRAM, one question comes to our mind naturally: can we achieve bulk SOT in a system with PMA which does not possess any well-marked global or local symmetry breaking?

The experiments below demonstrate the bulk SOT switching of amorphous CoTb single layer films. CoTb is chosen for the following reasons: (i) The observation of DMI and chiral domain walls in rare earth-transition metal (RE-TM) alloys provides a great possibility to realize bulk SOT; (ii) CoTb films have a large bulk PMA, and can be prepared by magnetron sputtering with industry compatibility; (iii) As a ferrimagnet, the magnetization of Co and Tb couple antiferromagnetically and the resultant net moment is significantly small, producing weak stray field and resultant high-density information storage; (iv) The composition is widely adjustable, providing more chance to study the mechanism of bulk SOT, which is limited by single crystal systems; (v) The amorphous CoTb has robust interfacial energies for the high quality growth of MgO barrier, providing large tunneling magnetoresistance for junction integration [21,22].

A series of SiN (3 nm)/$Co_{100-x}Tb_x$ ($t$ nm)/SiN (3 nm) films with PMA were prepared by sputtering on thermally oxidized Si substrate. Besides all of $Co_{100-x}Tb_x$ films with $x = 39$ and 34, $Co_{70}Tb_{30}$, $Co_{75}Tb_{25}$ and $Co_{80}Tb_{20}$ samples were provided by Prof. W. J. Jiang' group. 3 nm-thick SiN was employed as both buffer and capping layer to avoid any oxidization of CoTb and provide completely symmetric upper and lower interfaces to exclude the Rashba effect. We first checked the transport properties of 6 nm-thick $Co_{61}Tb_{39}$ films with the eight-terminal device whose channel width is 5 μm [Fig. 1(a)]. The anomalous Hall effect (AHE) were measured at two different current [Fig. 1(b)]. In CoTb alloys, since the 4$f$ band of Tb is far below the Fermi level, the anomalous Hall signal is dominated by the Co moment [23]. Therefore, the positive and negative values of anomalous Hall resistance ($\Delta R_H$) correspond to the Co moment pointing upward and downward (for the standard Hall measurement configuration), respectively, i.e. $\Delta R_H$ reflects the Co moment. Considering that the direction of the net moment is determined by z-axis magnetic field ($H_z$), the magnetization state can be easily recognized in Fig. 1(b). Since the net magnetic moment can be either parallel or anti-parallel with the Co moment at different temperature, the sign of AHE signal can be totally opposite. For $Co_{61}Tb_{39}$, when the applied current ($I$) is as low as 0.1 mA, the magnetic moment of Tb is dominant at room temperature, which is consistent with previous studies [24]. Noteworthily, $\Delta R_H$ changes sign at 2.5 mA, suggesting that the sample underdoes a transition from Tb dominant to Co dominant case due to the current heating. Here the reduction of $\Delta R_H$ is due to the reduction of Co magnetization at high temperature and square hysteresis loop is not observed because the temperature increase is not uniform throughout the channel.

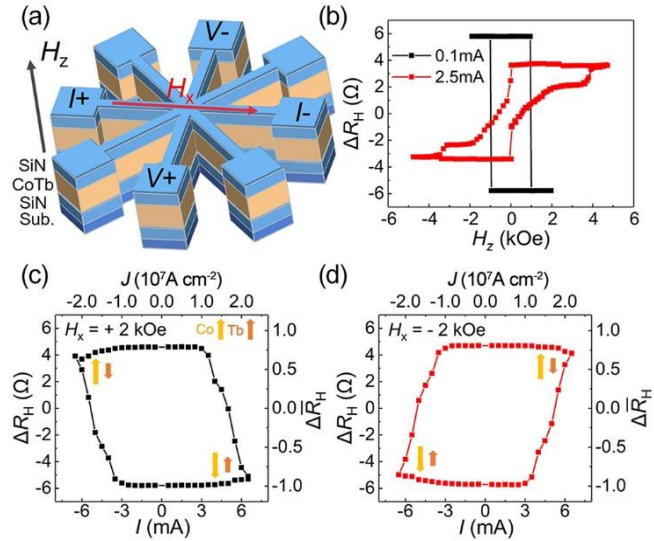

FIG. 1. (a) Sample layout and the schematic for transport measurements of the eight-terminal devices. (b) Anonalous Hall resistance ($\Delta R_H$) of $Co_{61}Tb_{39}$ with the applied current of 0.1 mA (blak curve) and 2.5 mA (red curve). Current-induced switching with the assitant field of (c) +2 kOe and (d) –2 kOe. The arrows with different colors denote the magnetization states of Co and Tb, and their length represents the magnitude of respective magnetic moment. The top axis shows the calculated current density of the 5 μm part of channel. The right axis shows the normalized anomalous Hall resistance ($\Delta \overline{R}_H$), which is normalized by the AHE under 0.1 mA.

Figures 1(c) and 1(d) present the SOT switching loop of $Co_{61}Tb_{39}$ samples under $H_x = +2$ kOe and $H_x = -2$ kOe, respectively. For this experiment, an assistant field ($H_x$) was applied along the current direction during scanning pulse current with 1 ms width which avoids the heating damage as much as possible and does not make the switching too difficult at the same time. $\Delta R_H$ was recorded after each pulse with a reading current of 0.1 mA. The opposite switching polarity under positive and negative $H_x$ is a strong evidence for the typical SOT switching, which demonstrates the bulk SOT in amorphous CoTb single layer films. Further analysis utilizing magneto-optical Kerr effect (MOKE) microscope shows that the switching is accomplished by the domain nucleation and expansion process [25]. To understand each data point in the SOT switching loop, a key factor is that the magnetization state of CoTb can be different under a large writing current and a small reading current, while the magnetization direction of Co, which can always be recognized through the sign of $\Delta R_H$, will keep unchanged during the cooling (from writing to reading) process. Considering that the large current pulse has transformed the sample to the Co dominant case, we can tell the detailed magnetization state during SOT switching, as displayed in Figs. 1(c) and 1(d). Such a switching polarity is the same as the classic substrate/Pt/CoFeB/MgO scenario [26], therefore we define the bulk spin Hall angle of $Co_{61}Tb_{39}$ to be positive. The normalized anomalous Hall resistance ($\Delta \overline{R}_H$) is also shown on the right axis, which is normalized by the anomalous Hall resistance range at 0.1 mA. During the current scanning process, the large current will heat the sample and cause the reduction of anomalous Hall resistance, like in Fig. 1(b). However, here $\Delta R_H$ is recorded 1.5 s after the large writing current with a reading current of 0.1 mA, where the sample has cooled down and the magnetization state has recovered. Therefore, the recorded $\Delta R_H$ can be compared with AHE measured at 0.1 mA, making this normalization applicable. The comparable Hall resistance induced by the current and magnetic field indicates the fully SOT switching. We also prepared (Co 0.3 nm/Tb 0.55 nm)$_7$ multilayers, which have the same atomic percentage of Tb with that of $Co_{61}Tb_{39}$ alloy, and observed analogical SOT switching behaviors [25]. In addition, by adding a 4 nm-thick Ta layer below or Pt layer above the $Co_{61}Tb_{39}$ layer, the switching polarity becomes opposite [25], which is consistent to previous SOT studies on the bilayers [9,24].

To understand the switching process of $Co_{61}Tb_{39}$ single layer films, we utilized the MOKE microscope to image the evolution of magnetic domains for a 20 μm-wide Hall bar. Corresponding SOT switching loops can be found in Supplemental Material [25]. With $H_x$ = +2 kOe, an inverse domain nucleates at the roughly central part of the current channel, and expand with increasing current from +11.5 to +12.3 mA, as displayed in Figs. 2(a)–2(e). An opposite switching can be also observed with increasing negative current [Figs. 2(f)–2(j)]. The switching for positive and negative current is slightly asymmetric mainly because $H_x$ is not perfectly in-plane. The critical switching current density is estimated to be ~1 × 10$^7$ A cm$^{-2}$, which is comparable to the SOT switching in heavy metal/ferromagnet systems [1,6,9,15]. Unexpectedly, the fully switching can occur with the assistant field as low as +20 Oe [Figs. 2(k)–2(m)], while in heavy metal/CoTb systems a large assistant field is usually needed to overcome the high interfacial DMI field [24]. Further analysis with the current-induced zero field "half-switching" of the magnetization [Figs. 2(n) and 2(o)] implies that in $Co_{61}Tb_{39}$ single layer films, the Oersted field plays a pivotal role in the SOT switching when the assistant field is low [25].

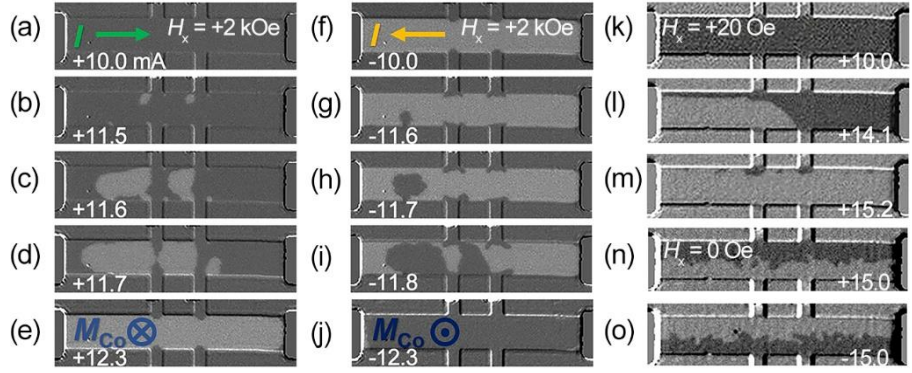

FIG. 2. Magnetization evolution of the 20 μm-wide Hall bar under the assistant field of +2 kOe with increasing (a-e) positive current and (f-j) negative current. (k-m) Fully SOT switching with $H_x$ = +20 Oe. (n,o) Current-induced zero field "half-switching" of the magnetization, which is actually the effect of the Oersted field. The direction of current and magnetic moment of Co (because only Co moment contributes to the MOKE signal) are marked in detail.

A series of $Co_{61}Tb_{39}$ films ($t$ = 4, 8, 10, and 15 nm) were also prepared and identical SOT switching measurements were carried out. Fully switching is achieved in all of the samples, and the switching polarity remains the same with the 6 nm-thick one. To roughly understand the influence of the film thickness on the SOT switching, we extract the critical switching current ($I_{SW}$), where half change of $\Delta R_H$ is detected. The assistant field keeps at +2 kOe for all of the samples to make sure that the switching efficiency has saturated, e.g. further increasing the assistant field would not decrease $I_{SW}$ anymore. $I_{SW}$ as a function of film thickness is presented in Fig. 3(a). Remarkably, $I_{SW}$ increases approximately linearly as $t$ increases, revealing that the critical switching current density keeps almost constant. That is, the switching does not become more difficult for a thicker $Co_{61}Tb_{39}$ film, which clearly excludes the interfacial origin and clarifies a bulk characteristic of the SOT switching.

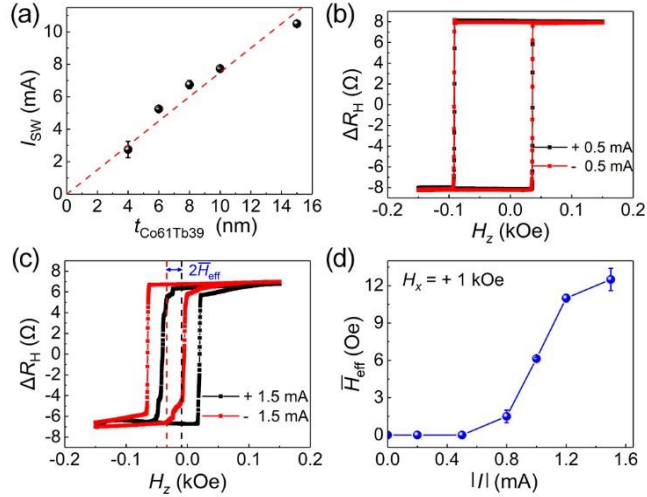

FIG. 3. (a) Critical switching current for SOT switching versus the thickness of $Co_{61}Tb_{39}$ films. The red dashed line is a linear fit. Anomalous Hall curves for dc current of (b) ±0.5 mA and (c) ±1.5 mA with $H_x$ = +1 kOe. The bias of the whole loop is because $H_x$ is not exactly in the sample plane. The black (red) dashed line in (c) is a schematic of the central position of the AHE curve for +1.5 mA (–1.5 mA). (d) $\overline{H}_{eff}$ versus the absolute value of current with $H_x$ = +1 kOe. The error bars in (a) and (d) are estimated from the s.d. of 3 measurements.

We now turn towards the estimation of the SOT effective field in $Co_{61}Tb_{39}$ single layer films. Current-induced switching loop shift [26] is used to estimate the effective field of the 4 nm-thick $Co_{61}Tb_{39}$ sample. This sample is chosen for its relatively small coercivity due to the relatively large interfacial influence for the small thickness and resultant weak PMA, which can shorten the measurement time and reduce the thermal damage to the greatest extent. The harmonic Hall measurements are not convenient for quantitative analysis due to the large anomalous Nernst effect of the CoTb single layer [25,27]. We recorded $\Delta R_H$ during scanning $H_z$ with an external filed of $H_x$ = +1 kOe applied along the (positive) current direction. As illustrated in Figs. 3(b) and 3(c), AHE loops of +0.5 mA and –0.5 mA do not exhibit any observable shift, while a clear shift of 25 Oe identifies the existence of SOT effective field when the current increases to 1.5 mA. By averaging the shift for positive and negative current, we could get the average SOT effective field $\overline{H}_{eff}$ and corresponding $\overline{H}_{eff}$ for different current, as plotted in Fig. 3(d). No obvious $\overline{H}_{eff}$ appears when the current is smaller than 0.5 mA. This threshold may be caused by the pinning of the domains in the device. Different from traditional heavy metal/ferro(ferri)magnet cases, the present $\overline{H}_{eff}$ shows a nonlinear increase with increasing current, which is possiblely due to the heat-related magnetism variation because both the saturation magnetization and magnetic anisotropy are sensitive to the temperature taking the sustained flow of the dc current during the measurement into account. The SOT effective field is at the same order of magnitude as that of classic Pt(Ta)/CoFeB/MgO system [26], indicating an efficient SOT swithcing in $Co_{61}Tb_{39}$ single layer.

We then address the question how the Tb-concentration affects SOT switching in CoTb single layer films. Four 6 nm-thick CoTb films were prepared with the composition of $Co_{66}Tb_{34}$, $Co_{70}Tb_{30}$, $Co_{75}Tb_{25}$ and $Co_{80}Tb_{20}$, in which at room temperature the former two and the latter two samples exhibit Tb-dominant and Co-dominant magnetization, respectively [25]. Figure 4 shows the SOT switching curves of this series of CoTb samples with positive assistant fields. The SOT switching of $Co_{66}Tb_{34}$ is similar to that of $Co_{61}Tb_{39}$, showing a clockwise switching loop in Fig. 4(a). Differently, $Co_{70}Tb_{30}$ exhibits a counterclockwise switching loop in Fig. 4(b). In contrast to the fully switching of $Co_{61}Tb_{39}$, $\Delta \overline{R}_H$ of $Co_{70}Tb_{30}$ is reduced to the scale of –1~0, indicating that only ~50% of domains can be switched within the current limit. Note that both $Co_{66}Tb_{34}$ and $Co_{70}Tb_{30}$ are switched above the compensation point [25], where the Co moment dominates the net moment. Therefore, opposite bulk spin Hall angle of these two samples can be deduced. Moreover, a higher assistant field of 5 kOe is needed to observe the most efficient SOT switching of the $Co_{70}Tb_{30}$ sample, and the switching is quite sensitive to the assistant field direction. Once the assistant field is slightly out of the sample plane, a preferred magnetization state would exist and a clear SOT switching loop is hardly observed, which means that the SOT effective field is relatively small and the external field plays a dominant role on the magnetization state. This is supported by identical SOT switching features observed in (Co 0.3 nm/Tb 0.4 nm)$_7$ multilayers with a comparable composition [25].

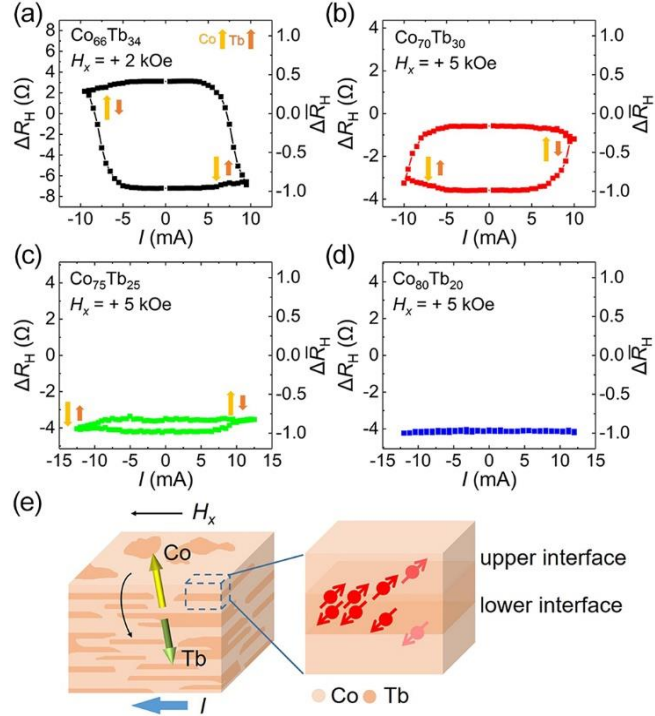

FIG. 4. SOT switching of (a) $Co_{66}Tb_{34}$, (b) $Co_{70}Tb_{30}$, (c) $Co_{75}Tb_{25}$ and (d) $Co_{80}Tb_{20}$ under positive assistant fields. The corresponding magnetization state of Co and Tb are represented by pale and deep yellow arrows, respectively. (e) Schematic of the spin current generation and transport. The left part illustrates the quasi-ordered state with many inner interfaces inside the CoTb alloy. SHE would generate equal spins with opposite polarization, however the transparency of the upper and lower inner interfaces is not the same, which is schematic by the different colors of the two interfaces in the right part.

Further reducing the Tb-composition to 25 would not change the sign of bulk spin Hall angle compared with the $Co_{70}Tb_{30}$ case, whereas the switching becomes more difficult and only ~10% of domains could be switched, as displayed in Fig. 4(c). For the most Co-rich $Co_{80}Tb_{20}$ sample, there is no trace of SOT switching in Fig. 4(d) anymore. (Co 0.32 nm/Tb 0.34 nm)$_5$ multilayers with comparably low Tb concentration also show negligible SOT effective field [9]. These composition-dependent SOT switching data clearly verify that SOT effective field depends strongly on the Tb concentration. When the Tb concentration decreases, the SOT effective field is reduced and undergoes a sign change. This will be further discussed below.

Ideally, if our substrate/SiN/CoTb/SiN system is considered as uniform and completely disordered amorphous films, no deterministic current-induced magnetization switching can be realized according to the symmetry-based analysis. However, the bulk PMA of CoTb actually provides some clues to re-understand the degree of order in this amorphous material. Previous studies have confirmed that the magnetic anisotropy in RE-TM alloy is relevant to the structural anisotropy by using extended x-ray-absorption fine structure characterizations, and the PMA can be developed when more RE-TM near neighbor pairs appears in the out-of-plane direction [28,29]. Therefore, the present CoTb alloy with PMA indicates more Co-Tb bonding exists in the out-of-plane direction rather than the in-plane direction, which creates many inner interfaces naturally by the perpendicular Co-Tb bonding [illustrated in Fig. 4(e)]. In this way, we could treat perpendicularly magnetized CoTb as a quasi-ordered material, which is similar to the existence of different types of ordering in various amorphous materials [30]. Because the films possess a fixed growth direction from bottom to top, it is natural to generate different chemical environments between lower and upper interfaces, e.g. the strain of Co/Pd and Pd/Co interfaces are largely different in Co/Pd multilayer [20], and in Cu/W multilayer the W on Cu interfaces are relatively sharp while the Cu on W interfaces are diffuse [31]. Therefore, the inner interfaces in CoTb would create asymmetry inside the films.

The existence of symmetry breaking indeed permits the deterministic SOT switching, while the spin current source is still under exploration. AHE can be excluded because the switching is independent with the magnetization direction [32]. The most dramatic feature in our results is that the sign of SOT effective field can be modulated by the composition. This sign change of effective filed reminds us the sign change of spin Hall angle in 4$d$ and 5$d$ transition metals, which is related to the number of $d$ electrons [33]. Magnetic materials could also have sizable spin Hall effect (SHE), such as FePt [34]. We propose that in CoTb, both 3$d$ electron of Co and the 4$f$ electron of Tb contribute to the SHE which could influence the sign of the spin Hall angle with the change of the composition, while the heavy element Tb with a stronger spin-orbit coupling has a major influence on the strength of the SHE. The SOT switching could be realized by the combination of SHE and asymmetric transport of spin current for upper and lower inner interfaces [illustrated in Fig. 4(e)]: spin current is generated uniformly in the bulk CoTb single layer, producing equal spins with opposite polarization; however, the penetrability of spin travelling upward and downward is different due to the asymmetry of the two kinds of inner interfaces, and the local magnetization would feel a net spin polarization. This process would emerge everywhere inside the films, producing SOT switching with a bulk character. It is worth pointing out that the present bulk SOT switching cannot be explained by the composition gradient along the growth direction in CoTb films: (i) it is difficult to image the reversal of composition gradient and the concomitant sign change of SOT just by changing Tb-concentration; (ii) our Co/Tb multilayers with negligible composition gradient also show clear bulk SOT switching.

Combining a sizable SHE which provides spin polarization and an out-of-plane ordering which creates asymmetry by inner interfaces, bulk SOT switching can be generalized to more thin film systems. A fully SOT switching in a nominal symmetric [(Co 0.4 nm/Pd 0.8 nm)$_2$/Co 0.4 nm] multilayer is observed [25]. $L1_0$-ordered FePt single layer was reported to possess SOT switching as well [35]. In contrast, we observe no current-induced switching in a 10 nm-thick Heusler alloy Co$_2$MnAl thin film with strong PMA [36], which should be attributed to the relative weak SHE. Although the precisely quantitative analysis is difficult at present due to the complex dependence of magnetism of CoTb on the film thickness and sample temperature [25], the thickness and composition dependent SOT switching make it straightforward to the proposed mechanism. Moreover, the inner interfaces and the ordering of CoTb are tunable during growing and processing. For instance, the PMA of alloy can be lost if depositing the film at ultra-high base vacuum or giving annealing treatment [28], which means that the inner interfaces are removed and the material becomes totally disordered. Or for the multilayer sample, the thickness for Co and Tb layer can be controlled to change the number of interfaces. We believe the rich structural modulation possibility of CoTb alloy and multilayer can help better understand the bulk SOT mechanism in future studies.

In summary, we demonstrate the efficient bulk SOT switching in CoTb single layer films. Careful analysis of the composition-dependent SOT switching implies that a combination of SHE and the asymmetric spin current transport can be the underneath mechanism, which could be generalized to produce bulk SOT switching in thin films with no well-marked global or local inversion broken symmetry but possessing a sizable SHE plus an out-of-plane ordering. Our findings not only open a window for the bulk SOT switching, which is benefit for the thermal stable SOT-MRAM, but also represent a significant step towards the amorphous structure exploitation from the symmetry breaking.

We acknowledge Prof. Jianhua Zhao and Dr. Zhifeng Yu for providing $Co_2MnAl$ thin films as well as the fruitful discussions with Prof. Dazhi Hou and Prof. Yang Shao. C.S. acknowledeges the support of Beijing Innovation Center for Future Chip (ICFC), Tsinghua University. This work is supported by the National Key R&D Program of China (Grant No. 2017YFB0405704), National Natural Science Foundation of China (Grant Nos. 51871130 and 51671110), and National Key R&D Program of China (Grant Nos. 2016YFA0203800 and 2017YFB0405604).